\documentclass[10pt,journal]{IEEEtran}

\usepackage{cite}
\usepackage{graphicx}
\usepackage{algorithm}
\usepackage{algorithmic}
\usepackage[hyphens]{url}
\usepackage{amsmath}
\usepackage{wrapfig}

\newtheorem{lemma}{Lemma}

\renewcommand{\algorithmicrequire}{\textbf{Input:}}
\renewcommand{\algorithmicensure}{\textbf{Output:}}

\begin{document}

\title{Joint Link Scheduling and Brightness Control for Greening VLC-based Indoor Access Networks}

\author{Sihua Shao, Abdallah Khreishah, Issa Khalil
\thanks{Sihua Shao and Abdallah Khreishah are with the Department of Electrical and Computer Engineering,
New Jersey Institute of Technology, email: ss2536@njit.edu, abdallah@njit.edu}
\thanks{Issa Khalil is with the Qatar Computing Research Institute and Qatar Foundation, email: ikhalil@qf.org.qa}\vspace{-20pt}}

\maketitle

\begin{abstract}
Demands for broadband wireless access services is expected to outstrip the spectrum capacity in the near-term - ``spectrum crunch". Deploying additional femotocells to address this challenge is cost-inefficient, due to the backhaul challenge and the exorbitant system maintenance. According to an Alcatel-Lucent report, most of the mobile Internet access traffic happens indoor. Leveraging power line communication and the available indoor infrastructure, visible light communication (VLC) can be utilized with small one-time cost. VLC also facilitates the great advantage of being able to jointly perform illumination and communications, and little extra power beyond illumination is required to empower communications, thus rendering wireless access with small power consumption. In this study, we investigate the problem of minimizing total power consumption of a general multi-user VLC indoor network while satisfying users' traffic demands and maintaining an acceptable level of illumination. We utilize the column generation method to obtain an $\epsilon$-bounded solution. Several practical implementation issues are integrated with the proposed algorithm, including different configurations of light source and ways of resolving the interference among VLC links. Through extensive simulations, we show that our approach reduces the power consumption of the state-of-art VLC-based scheduling algorithms by more than 60\% while maintaining the required illumination.

\end{abstract}

\section{Introduction}\label{introduction}
The Internet has emerged as an integral part of our lives, which has gone far beyond its original use for connecting computers, and now also inter-connects mobile phones, devices, and everything. The traffic from wireless and mobile devices will exceed that from wired devices by 2019 \cite{cisco2015cisco}. By the same year, WiFi and mobile devices will account for 66\% of IP traffic. According to \cite{Ericsson2013report}, we spend 90\% of our time indoor and 80\% of the mobile Internet access traffic happens indoor \cite{Alcatel2015inbuilding,Cisco2015service}. This percentage will only increase as 54\% of the cellular traffic is expected to be offloaded to WiFi by 2019 \cite{index2015global}. It is expected that 87\% of the companies would switch providers by 2019 for better indoor coverage \cite{Alcatel2015inbuilding}.


Relying on cellular networks alone to satisfy such demand is not a viable solution, because the size of the cell can not be decreased in an arbitrary fashion due to the backhaul challenge \cite{chia2009next} and the increased cost of building and maintaining the system. This fact puts pressure on indoor networks to support the insatiable demand for wireless Internet access.


To alleviate the problems of system construction and maintenance, and to enhance spectrum reusability, innovative approaches need to be adopted, among which visible light communication (VLC) is an excellent candidate. The emergence and the commercialization of power line communication concept \cite{tsuzuki2012feasibility,tonello2008challenges}, makes it easy and attractive to add a driver circuit to perform modulation functionality between the light source and the power cables and utilize VLC with very small one-time cost based on the available indoor infrastructure. Benefiting from the short transmission range and unlicensed wide bandwidth, VLC technology provided with LED devices is characterized by high area spectral efficiency. As a complementary approach to the existing wireless RF solutions, VLC is poised to overcome the crowded radio spectrum and become a promising broadband wireless access candidate to resolve the ``spectrum crunch" problem \cite{hanchard2010fcc}.

For indoors environment, whenever communication is needed, {\it lighting is also needed most of the time}. According to \cite{lindsay2012what}, energy consumption of lighting represents about 15\% of the worlds total energy consumption. Therefore, by jointly performing lighting and Internet access, VLC can operate on a very small energy budget.


For upper layers in multi-user scenarios, \cite{li2012vico} and \cite{tao2015scheduling} consider link scheduling algorithms to serve multiple users based on multiple VLC links with no power efficiency guarantees. The work in \cite{li2012vico} relies on a simple impractical assumption to measure the illumination using the average SNR distribution. The work in \cite{din2014energy} proposes an energy-efficient brightness control and data transmission scheme for VLC. However, the scheme can only be applied to the single user scenario and only the optical power is taken into account when minimizing the power consumption.

In this paper, we investigate the problem of {\it optimizing total power consumption of a general multi-user VLC indoor network while satisfying the traffic demands and illumination requirements}. A novel algorithm is proposed to efficiently obtain a practical $\epsilon$-bounded solution. Our contributions are summarized as follows:

\begin{itemize}
\item {\bf Minimizing the total power consumption for a general multi-user VLC indoor network:} Taking the users' traffic demand and the illumination requirement of entire horizontal space into account, the total power consumption of a general multi-user VLC indoor network is optimized via a novel efficient and practical algorithm.
\item {\bf Effectively model the level of interference among VLC links:} Based on the proposed algorithm, an effective interference management approach is verified by extensive simulation results.
\item {\bf Design of a novel structure of light source:} An innovative configuration of light source is proposed and analyzed, and compared with two other common configurations in terms of total power consumption.
\item {\bf Validating the power efficiency and illumination satisfaction:} Extensive simulation results reveal that, our proposed link scheduling algorithm can provide around 60\% and 80\% saving in power consumption compared to two VLC-based solutions, and the illumination distribution obtained by our proposed algorithm can always satisfy the requirements.
\end{itemize}


\section{System Model}\label{system_model}
\subsection{Access system model}
\begin{figure}
\centering
\includegraphics[width=0.37\textwidth]{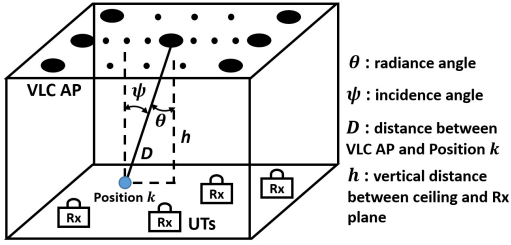}
\caption{System model for multi-user VLC indoor network}
\vspace{-12pt}
\label{fig_system_model}
\end{figure}

Consider a visible light access system model (Fig.~\ref{fig_system_model}) comprising of $\mathcal{V}=\{1,2,...,i,...,V\}$ VLC APs, $\mathcal{M}=\{1,2,...,j,...,M\}$ user terminals (UTs), and $\mathcal{W}=\{1,2,...,b,...,W\}$ available channels with different bandwidths. The bandwidth of channel $b$ is denoted by $B^{b}$. We also denote the set of transmitters on VLC AP $i\in\mathcal{V}$ by $Tx_{i}=\{1,2,...,m,...,|Tx_{i}|\}$, where $|Tx_{i}|$ is the number of transmitters on VLC AP $i$, and the set of receivers on UT $j\in\mathcal{M}$ by $Rx_{j}=\{1,2,...,n,...,|Rx_{j}|\}$, where $|Rx_{j}|$ is the number of receivers on UT $j$. We assume the UT $j$ has a throughput requirement $\mathcal{R}_{j}$.

A recent measurement study \cite{ding2013characterizing} on traces of 3785 smart phone users from 145 countries over a four-month period shows that the ratio of WiFi download traffic to its upload traffic is 20:1. Therefore, in this work, we mainly consider the power consumption for the downlink data transmission. Regarding the uplink issue in VLC network, a hybrid WiFi-VLC Internet access system (VLC downlink and WiFi uplink) is presented in our earlier work \cite{shao2014indoor,shao2015analysis,shao2015design}.

\subsection{Communication}
Optical modulation is performed by varying the forward current of the light source. The output optical power changes proportionally to the modulated forward current. The increase in total power consumption (including the power consumed by modulator) is mainly due to the switching loss in the driver circuitry at high speed (AC current for modulation). Such behavior is observed in our preliminary experimental results and the results in \cite{hsu2005optimization}. Here, we denote a peak-to-peak optical signal strength (generated from AC current) by $P_{AC}$ and its average value by $P_{AC,Avg}$.

\subsection{Illuminance}
Consider a horizontal user plane comprising of $\mathcal{K}=\{1,2,...,k,...,K\}$ positions and each position requires an illumination level in the range of $E_{k}^{L}$ and $E_{k}^{U}$. The illumination level of ambient light at position $k$ is denoted by $E_{k}^{Am}$.

The illumination level at a given location depends on the average optical power received. This can be generated by both the DC and the AC current. We denote an optical DC power (generated from DC current) by $P_{DC}$, which is responsible for compensating the average AC power in order to meet the illumination demands.

The DC component does not require a current switching process. This switching process reduces the efficiency of the driver circuit and light source by consuming more power. Thus, for the transmitter $m$ of the $i^{th}$ VLC AP, we denote $\eta_{AC}^{i,m}$ and $\eta_{DC}^{i,m}$ as the wall plug efficiency factors (i.e. the ratio of the optical power consumption to the electrical power consumption) for AC and DC optical power, respectively, where $\eta_{AC}^{i,m}$ is generally smaller than $\eta_{DC}^{i,m}$.

The illuminance represents the level of brightness of the illuminated surface. A horizontal illuminance $E_{k}$ [lux] at position $k$, can be given \cite{din2014energy} as $E_{k}=\sum_{i\in\mathcal{V}}\sum_{m\in Tx_{i}}(P_{DC}^{i,m}g_{i,k,m}^{DC}+P_{AC,Avg}^{i,m}g_{i,k,m}^{AC})\rho$, where $P_{DC}^{i,m}$ and $P_{AC,Avg}^{i,m}$ denote the generated DC and average AC optical power from the transmitter $m$ of the $i^{th}$ VLC AP, respectively, $\rho$ [lm/W] is the luminosity efficacy, and $g_{i,k,m}^{DC}$ and $g_{i,k,m}^{AC}$ are given as follows
\begin{align}
g_{i,k,m}^{DC}&=\frac{ml_{DC}+1}{2\pi D^{2}}cos^{ml_{DC}}(\theta_{DC})cos(\psi)\nonumber\\
g_{i,k,m}^{AC}&=\frac{ml_{AC}+1}{2\pi D^{2}}cos^{ml_{AC}}(\theta_{AC})cos(\psi)\nonumber
\end{align}
where $ml$ is the Lambertian order ($ml=-ln2/ln(cos\theta_{1/2})$, $\theta_{1/2}$ is the semi-angle at half power), (in Fig.~\ref{fig_system_model}) $D$ is the distance between VLC AP $i$ and position $k$, $\theta$ is the radiance angle, and $\psi$ is the incidence angle. Note that, the $ml$ and $\theta$ of the DC and AC powered source could be different, which will be discussed in detail in Section~\ref{pracitcal_issues}.

\subsection{Channel capacity, interference and noise}
From \cite{komine2004fundamental}, the received optical intensity via a line-of-sight (LOS) path is about 30 times higher than that via the first reflective path. Since LOS paths are typically feasible, in this paper, we only consider the LOS links. The LOS channel gain between VLC AP $i$, using transmitter $m$, and UT $j$, using receiver $n$, denoted by $H_{ij,mn}$, is illustrated in \cite{komine2004fundamental}.

It is worth noticing that, the gain of VLC channel highly depends on the strict alignment between VLC transceivers. In contrast to the omnidirectional WiFi channel, the potential motion of users will lead to severe degradation of VLC channel gain. Nevertheless, according to \cite{Cisco2015service}, most of the Internet access traffic will happen indoor at fixed locations. Thus, we focus on scenarios that the users' location are fixed, where the channel gain of VLC is more stable than that of WiFi \cite{zhang2015dancing}.

The Gaussian noise of the optical wireless channel consists of the shot noise (stems from the received optical power) and the thermal noise (stems from the receiver's circuitry). Increasing the transmitted optical power increases the noise level at the receiver. However, if the modulation bandwidth is large (above 50 MHz) and the optical power level is low (below 20 W) as is the case with most VLC APs and VLC front-ends \cite{langer2013optoelectronics}, the thermal noise would dominate the shot noise \cite{komine2004fundamental}. With fixed gain of the receiver, the thermal noise is essentially independent of the ambient light and signal strength while the shot noise is not. Our experimental results \cite{shao2015analysis,shao2015design} have validated this behavior even in outdoor settings. Therefore, a constant variance of Gaussian noise can be assumed.

Typically, there are two major interference models for the wireless networks \cite{gupta2000capacity}: Physical Model and Protocol Model. Under Physical Model, the link capacity depends on the signal-to-interference-plus-noise ratio (SINR) at the receiver side. It is a very accurate representation of real scenarios, but it is computationally difficult to work with. On the other hand, Protocol Model only considers the pairwise interference relationship among the links. With Protocol Model, the interference among individual VLC links can be modeled through the use of a conflict graph \cite{diestel2005graph}. The approach works as follows. Each receiver on a UT or each transmitter on a VLC AP is represented by a vertex in a graph. If a transmitter can transmit to a receiver on a given channel, an edge is drawn between the two vertices representing the transmitter and the receiver. The conflict graph is then constructed, such that each edge in the original graph is represented by a vertex in the conflict graph. An edge in the conflict graph is drawn between two vertices, if the corresponding edges in the original graph interfere with each other. An interference constraint (addressed in details in the next section) represents the fact that for a successful transmission (i.e. vertex in the conflict graph), none of those vertices (i.e. links in the original graph) connected by an edge in the conflict graph are active at the same time. Based on this interference constraint, the link capacity depends on the signal-to-noise ratio (SNR) under Protocol Model.


Given a link with bandwidth $B^{b}$, based on Shannon-Hartley theorem \cite{cover2012elements}, the maximum link capacity $C$ when VLC AP $i$, using transmitter $m$, transmits data to UT $j$, using receiver $n$, on channel $b$, in two interference models are given \cite{stefan2014hybrid} by
\begin{align}
&C_{ij,mn}^{b}(Protocol)=B^{b}\times log_{2}(1+\frac{(\gamma H_{ij,mn}P_{AC}^{i,m})^{2}}{N})\\
&C_{ij,mn}^{b}(Physical)=B^{b}\times log_{2}(1+\frac{(\gamma H_{ij,mn}P_{AC}^{i,m})^{2}}{(\gamma P_{I})^{2}+N})
\end{align}
where $\gamma$ is the detector responsivity, $P_{I}$ is the summation of interference optical power and $N$ is the variance of noise.

\section{Problem Formulation}\label{problem_formulation}
We investigate the minimum power consumption problem for a multi-user VLC indoor network by joint link scheduling and illuminating. For a conflict graph under Protocol Model, an {\it independent set} (IS) $\mathcal{I}$ is defined as a set of vertices in the conflict graph (i.e. links in the original graph) such that none of them are connected by an edge \cite{diestel2005graph}.


Suppose all the ISs are known and the set of all ISs is denoted as $\mathcal{Q}=\{\mathcal{I}_{1},\mathcal{I}_{2},...,\mathcal{I}_{q},...,\mathcal{I}_{|\mathcal{Q}|}\}$. In order to ensure the successful transmissions in each IS, at any given time, only one IS should be active. We define $\omega_{q}$ as the fraction of the time during which the $q^{th}$ IS is active. Therefore, we have
\begin{align}
&\sum_{1\leq q\leq|\mathcal{Q}|}\omega_{q}\leq 1,~\omega_{q}\geq 0
\end{align}

An integer variable is defined as follows: $x_{ij,mn}^{q,b}$ is equal to 1 if VLC AP $i$, using transmitter $m$, transmits to UT $j$, using receiver $n$, on channel $b$, in $\mathcal{I}_{q}$, and equal to 0 otherwise.

Recall that the $j^{th}$ UT's throughput requirement is $\mathcal{R}_{j}$. In order to meet the traffic demands of users, the following set of constraints need to be satisfied
\begin{align}
\sum_{1\leq q\leq|\mathcal{Q}|}\omega_{q}\sum_{i\in\mathcal{V}}\sum_{b\in\mathcal{W}}\sum_{m\in Tx_{i}}\sum_{n\in Rx_{j}}\mathcal{C}_{ij,mn}^{b}x_{ij,mn}^{q,b}\geq \mathcal{R}_{j}\nonumber\\
(\forall j\in\mathcal{M})
\end{align}
where $\mathcal{C}_{ij,mn}^{b}$ is calculated by (1).

Denote $P_{DC}^{i,q,m}$ as the $i^{th}$ VLC AP's DC optical power consumption of the transmitter $m$ in the $q^{th}$ IS and $P_{AC}^{i,m}$ as the fixed peak-to-peak signal strength of $i^{th}$ VLC AP's transmitter $m$. For one VLC transmitter, the summation of $P_{AC}$ and $P_{DC}$ can not exceed the maximum optical power $P_{max}$. Thus we have
\begin{align}
P_{DC}^{i,q,m}+\sum_{j\in\mathcal{M}}\sum_{b\in\mathcal{W}}\sum_{n\in Rx_{j}}P_{AC}^{i,m}x_{ij,mn}^{q,b}\leq P_{max}^{i,m}\nonumber\\
(\forall i\in\mathcal{V}, \forall m\in Tx_{i}, 1\leq q\leq|\mathcal{Q}|)
\end{align}

Recall that, at position $k$, the minimum and maximum illuminance thresholds are $E_{k}^{L}$ and $E_{k}^{U}$, respectively, and the illuminance level of ambient light is $E_{k}^{Am}$. The summation of the LED lighting and the ambient lighting needs to be within the range [$E_{k}^{L}$,$E_{k}^{U}$]. Therefore, we have the following illumination constraints
\begin{align}
&E_{k}^{U}\geq\sum_{i\in\mathcal{V}}\sum_{j\in\mathcal{M}}\sum_{b\in\mathcal{W}}\sum_{m\in Tx_{i}}\sum_{n\in Rx_{j}}(P_{AC,Avg}^{i,m}x_{ij,mn}^{q,b}g_{i,k,m}^{AC,q}+\nonumber\\
&P_{DC}^{i,q,m}g_{i,k,m}^{DC})\rho+E_{k}^{Am}\geq E_{k}^{L}~(\forall k\in\mathcal{K}, 1\leq q\leq|\mathcal{Q}|)
\end{align}
The reason for adding the superscript $q$ to the AC optical power gain is that, the Lambertian order or the radiance angle of the AC powered source on each VLC transmitter may be varied in different IS. This condition will be demonstrated in details in Section~\ref{pracitcal_issues}.

Given the link scheduling in each IS, the illuminance distribution from the average AC optical power can be obtained. Therefore, to satisfy the maximum power constraint (5) and the illumination constraint (6), we can compute the optimal $P_{DC}^{i,q,m}$ for each VLC transmitter in each IS. Denote $P_{AC,Avg}(\mathcal{I}_{q})$ and $P_{DC}(\mathcal{I}_{q})$ as the total AC and DC power consumption of the $q^{th}$ IS, respectively. The optimal solution of our algorithm might result in $\sum_{1\leq q\leq|\mathcal{Q}|}\omega_{q}<1$, this means that the data transmission will be completed within the $\sum_{1\leq q\leq|\mathcal{Q}|}\omega_{q}$ fraction of time. However, the illumination is always needed. Let $P_{illumi}^{min}$ represent the minimum total power consumption when all the VLC APs only perform illumination. This means that during the $(1-\sum_{1\leq q\leq|\mathcal{Q}|}\omega_{q})$ fraction of time, the power consumption is $P_{illumi}^{min}$. Therefore, the total power consumption optimization problem can be formulated as follows

\begin{align}
\min_{\omega_{q}}~&\sum_{1\leq q\leq|\mathcal{Q}|}\omega_{q}[P_{AC,Avg}(\mathcal{I}_{q})+P_{DC}(\mathcal{I}_{q})]\nonumber\\
&\qquad\qquad\qquad\qquad+(1-\sum_{1\leq q\leq|\mathcal{Q}|}\omega_{q})P_{illumi}^{min}\nonumber\\
s.t.~&(3),(4)\nonumber
\end{align}

Given that all the ISs satisfying constraints (5) and (6), the formulated optimization problem is a linear programming problem. We call this problem the {\it master problem} (MP). The solution to the MP is to find the optimal values of $\omega_{q}$ ($1\leq q\leq|\mathcal{Q}|$). In the next section, we will introduce the challenges of solving the MP and our solution methodology.

\section{Solution Methodology}\label{solution_methodology}
\subsection{Challenges of Solving MP}
To efficiently solve the MP, there are two main challenges: i) Although the MP is a linear programming problem if all the ISs are given, the IS decision problem itself is NP-complete \cite{diestel2005graph} and hence it is believed that there is no efficient algorithm for solving it. ii) Even if all the ISs are given, the number of ISs and corresponding variables increases exponentially as the number of links increases. Therefore, the complexity of solving the MP will be extremely high when the network is very large. We propose a column generation based $\epsilon$-bounded approximation algorithm to resolve these challenges.
\subsection{Column Generation}
Column generation \cite{bertsimas1997introduction} is an efficient algorithm for solving large scale (i.e. the number of variables is large) linear programming problem. Even though the MP has a large number of variables, only a small subset of them will be non-zero (basis variables) in the optimal solution. Based on this observation, rather than adding all the variables in the MP, column generation only generates the variables with the highest potential to enhance the objective function. In particular, the large MP is split into two smaller and simpler problems: {\it restricted master problem} (RMP) and {\it pricing problem} (PP). The RMP only includes an initial subset of variables in the MP, and the PP is a new optimization problem assigned to find a variable or a column (i.e. independent set) that has the most negative reduced cost (i.e. decrease of the objective value). The process works iteratively as follows: the RMP is solved and its optimal and dual optimal solutions are obtained; the PP utilizes the dual optimal solution of the RMP to identify the column with the most negative reduced cost and adds it into RMP to re-optimize RMP. The process continues until the objective value of PP is non-negative. If the PP returns a non-negative solution, the solution of RMP is the optimal solution to the MP.

Instead of considering the set $\mathcal{Q}$ of all the ISs, RMP starts with an initial set $\tilde{\mathcal{Q}}$ of ISs (called observed ISs). A simple method of selecting the initial ISs is to place only one active link in each of them. Hence, the RMP is formulated as follows
\begin{align}
&\min_{\omega_{q}}~\sum_{1\leq q\leq|\tilde{\mathcal{Q}}|}\omega_{q}[P_{AC,Avg}(\mathcal{I}_{q})+P_{DC}(\mathcal{I}_{q})]\nonumber\\
& \qquad\qquad\qquad\qquad\qquad+(1-\sum_{1\leq q\leq|\tilde{\mathcal{Q}|}}\omega_{q})P_{illumi}^{min}\nonumber\\
s.t.~&\sum_{1\leq q\leq|\tilde{\mathcal{Q}}|}\omega_{q}\sum_{i\in\mathcal{V}}\sum_{b\in\mathcal{W}}\sum_{m\in Tx_{i}}\sum_{n\in Rx_{j}}\mathcal{C}_{ij,mn}^{b}x_{ij,mn}^{q,b}\geq R_{j}\nonumber\\
&\qquad\qquad\qquad \qquad\qquad\qquad\qquad\qquad\qquad (\forall j\in\mathcal{M})\nonumber\\
&\sum_{1\leq q\leq|\tilde{\mathcal{Q}}|}\omega_{q}\leq 1,~\omega_{q}\geq 0\nonumber
\end{align}
where $P_{DC}(\mathcal{I}_{q})$ ($1\leq q\leq|\tilde{\mathcal{Q}}|$) can be computed optimally by satisfying the maximum power constraint (5) and the illumination constraint (6).

After solving the RMP, the primal optimal solution and the Lagrangian dual optimal solution can be obtained. Since the ISs in RMP is only a subset of the ISs in MP (i.e. $\tilde{\mathcal{Q}}\subseteq\mathcal{Q}$), the primal optimal solution of RMP can be regarded as an upper bound of the optimal solution of MP. Adding another column, which does not exist in RMP, may reduce the upper bound and improve the objective function. Therefore, the PP is responsible for generating a column with the most negative reduced cost.

For a IS that has not been observed (i.e. included in the RMP), the PP needs to determine whether the reduced cost of the IS is negative or not. Referring to \cite{bertsimas1997introduction}, the reduced cost of $\mathcal{I}_{q}$ can be calculated as $c_{r}(\mathcal{I}_{q})-P_{illumi}^{min}$, where
\begin{align}
c_{r}(\mathcal{I}_{q})=\sum_{i\in\mathcal{V}}\sum_{j\in\mathcal{M}}\sum_{b\in\mathcal{W}}\sum_{m\in Tx_{i}}\sum_{n\in Rx_{j}}(\frac{1}{\eta_{AC}^{i,m}} P_{AC,Avg}^{i,m}x_{ij,mn}^{b}+\nonumber\\
\frac{1}{\eta_{DC}^{i,m}}P_{DC}^{i,m})-\sum_{j\in\mathcal{M}}\lambda_{j}\sum_{i\in\mathcal{V}}\sum_{b\in\mathcal{W}}\sum_{m\in Tx_{i}}\sum_{n\in Rx_{j}}\mathcal{C}_{ij,mn}x_{ij,mn}^{b}\nonumber
\end{align}
where $\lambda_{j}$ is the Lagrangian dual optimal solution for the $j^{th}$ UT. To find the $\mathcal{I}_{q}$ with the most negative reduced cost, the objective of PP is to minimize $c_{r}(\mathcal{I}_{q})-P_{illumi}^{min}$.

Regarding the constraints of PP, under Protocol Model, we denote $\mathcal{T}_{ij,mn}^{b}$ as the set of links that interfere with the transmission from the $i^{th}$ AP, using transmitter $m$, to the $j^{th}$ UT, using receiver $n$, on channel $b$. Thus, we have
\begin{align}
x_{ij,mn}^{b}+\sum_{pq,uv\in\mathcal{T}_{ij,mn}}x_{pq,uv}^{b}\leq 1\nonumber\\
(\forall i\in\mathcal{V}, \forall j\in\mathcal{M}, \forall m\in Tx_{i}, \forall n\in Rx_{j})
\end{align}
In addition, the total number of transmitting links at VLC AP $i$ and receiving links at UT $j$ should be no larger than $|Tx_{i}|$ and $|Rx_{j}|$, respectively, which means
\begin{align}
&\sum_{j\in\mathcal{M}}\sum_{b\in\mathcal{W}}\sum_{m\in Tx_{i}}\sum_{n\in Rx_{j}}x_{ij,mn}^{b}\leq |Tx_{i}|~(\forall i\in\mathcal{V})\\
&\sum_{i\in\mathcal{V}}\sum_{b\in\mathcal{W}}\sum_{m\in Tx_{i}}\sum_{n\in Rx_{j}}x_{ij,mn}^{b}\leq |Rx_{j}|~(\forall j\in\mathcal{M})
\end{align}

Besides, a transmitter of a VLC AP can not transmit to multiple UTs and a receiver of a UT can not receive from multiple VLC APs due to interference. Therefore, we get
\begin{align}
&\sum_{j\in\mathcal{M}}\sum_{b\in\mathcal{W}}\sum_{n\in Rx_{j}}x_{ij,mn}^{b}\leq1~(\forall i\in\mathcal{V}, \forall m\in Tx_{i})\\
&\sum_{i\in\mathcal{V}}\sum_{b\in\mathcal{W}}\sum_{m\in Tx_{i}}x_{ij,mn}^{b}\leq1~(\forall j\in\mathcal{M}, \forall n\in Rx_{j})
\end{align}

It is worth noticing that, in contrast to RF communication, the optical signal propagation may not be isotropic from the perspective of UTs plane (e.g. the central luminous flux is not vertical to the horizontal UTs plane), and also the orientation and filed of view (FOV) of VLC receiver can be tuned in order to receive signals from a specific direction and a small range. Thus, using the same channel, a transmitter of a VLC AP can transmit to multiple UTs, and a UT can receive from multiple VLC APs.

Assembling together the constraints of power and brightness control with the above constraints, the PP can be formulated as follows:
\begin{align}
&\min_{\mathcal{I}_{q}\in\mathcal{Q}\backslash\tilde{\mathcal{Q}}}c_{r}(\mathcal{I}_{q})-P_{illumi}^{min}\nonumber\\
s.t.~&(7),(8),(9),(10),(11)\nonumber\\
&P_{DC}^{i,m}+\sum_{j\in\mathcal{M}}\sum_{b\in\mathcal{W}}\sum_{n\in Rx_{j}}P_{AC}^{i,m}x_{ij,mn}^{b}\leq P_{max}^{i,m}\nonumber\\
&\qquad\qquad\qquad\qquad\qquad(\forall i\in\mathcal{V}, \forall m\in Tx_{i})\nonumber\\
&E_{k}^{U}\geq\sum_{i\in\mathcal{V}}\sum_{j\in\mathcal{M}}\sum_{b\in\mathcal{W}}\sum_{m\in Tx_{i}}\sum_{n\in Rx_{j}}(P_{DC}^{i,m}g_{i,k,m}^{DC}+\nonumber\\
&\qquad P_{AC,Avg}^{i,m}x_{ij,mn}^{b}g_{i,k,m}^{AC})\rho+E_{k}^{Am}\geq E_{k}^{L}~(\forall k\in\mathcal{K})\nonumber
\end{align}
where $P_{DC}^{i,m}$ and $x_{ij,mn}^{b}$ are the variables.

RMP and PP are solved in an iterative way, until PP returns a non-negative reduced cost. However, since the size of $\mathcal{Q}$ could be huge, it might take long time to reach the optimal solution. Therefore, instead of finding the optimal solution, we propose an $\epsilon$-bounded approximation approach to find a satisfactory $\epsilon$-bounded solution.

\subsection{$\epsilon$-Bounded Approximation Approach}
Let $z^{\ast}$ denote the optimal result of MP and $z^{u}$ denote the optimal result of RMP (upper bound on $z^{\ast}$), when $\kappa\geq\sum_{1\leq q\leq|\mathcal{Q}|}\omega_{q}$ holds for the optimal solution of MP, the $z^{u}$ can not be reduced more than $\kappa$ times the most negative reduced cost $c_{r}^{\ast}$ \cite{desrosiers2005primer} :
\begin{align}
&z^{u}+\kappa c_{r}^{\ast}\leq z^{\ast}\leq z^{u}\nonumber
\end{align}
Denote $z^{l}=z^{u}+\kappa c_{r}^{\ast}$ as the lower bound on $z^{\ast}$ and the following lemma can be proved.
\begin{lemma}
Define $\epsilon$ as $0\leq\epsilon<1$. The optimal solution of RMP $z^{u}\leq(1+\epsilon)z^{\ast}$, if $z^{u}/z^{l}\leq 1+\epsilon$.
\end{lemma}
\begin{IEEEproof}
If $z^{u}/z^{l}\leq 1+\epsilon$, then $z^{u}\leq(1+\epsilon)z^{l}\leq(1+\epsilon)z^{\ast}$. Hence, $z^{u}$ is the $\epsilon$-bounded solution of MP.
\end{IEEEproof}

Based on the $\epsilon$-bounded approximation approach, the iteration of column generation can be terminated when $c_{r}^{\ast}\geq0$ or $z^{u}/z^{l}\leq 1+\epsilon$.

\subsection{Reality Check}
Although the column generation based $\epsilon$-bounded approximation algorithm is capable of efficiently finding the $\epsilon$-bounded solution of MP, the feasibility of the Protocol Model solution is doubtful. Under the Protocol Model, the impact of some non-zero interfering links are neglected, which may lead to overestimation of the achievable link capacity. Thus the power consumption solution (under the Protocol Model) of MP may be lower than the real power consumption in practice. To investigate the practical objective value of MP, a validation process called ``reality check" is introduced in \cite{shi2013bridging}. Note that, in \cite{shi2013bridging}, the ``reality check" process is neither integrated with the column generation algorithm nor applied to evaluate the real power consumption.

Since the Protocol Model solution of RMP includes all the ISs with non-zero $\omega_{q}$, the link scheduling (i.e. which link is active and which link is idle) of each of those ISs can be known. Hence, the actual achievable capacity of each active link can be recomputed by equation (2). The ``reality check" process consists of two steps: i) Recompute the actual link capacity by equation (2) for the Protocol Model solution of RMP; ii) Substitute the recomputed link capacity into RMP and re-optimize the RMP to obtain a feasible solution of power consumption. The column generation based $\epsilon$-bounded approximation algorithm with reality check is summarized in Algorithm~\ref{algorithm}.


\begin{algorithm}
\caption{Column generation based $\epsilon$-bounded approximation algorithm with reality check}\label{algorithm}
\begin{algorithmic}[1]
\REQUIRE~Initial independent sets $\tilde{\mathcal{Q}}$, traffic demands $\mathcal{R}_{j}$, interference set of each link $\mathcal{T}^{b}_{ij,mn}$, Protocol Model link capacity $\mathcal{C}^{b}_{ij,mn}$, optical signal strength $P_{AC}^{i,m}$, maximum power consumption $P_{max}^{i,m}$, approximation factor $\epsilon$, gains of AC and DC optical power $g^{AC}_{i,k,m}$ and $g^{DC}_{i,k,m}$, wall plug efficiency factors $\eta_{AC}^{i,m}$ and $\eta_{DC}^{i,m}$, illuminance lower and upper bounds $E_{k}^{l}$ and $E_{k}^{u}$, ambient lighting level $E_{k}^{Am}$, $c_{r}^{\ast}=-\infty$, $\kappa=1$, $z^{l}=-\infty$ and $z^{u}=\infty$.\\
\ENSURE~Time fraction (out of one time unit) $\omega_{q}$ and solution of RMP $z^{u}$.\\
\STATE Compute $P_{AC,Avg}(\mathcal{I}_{q})$ and $P_{DC}(\mathcal{I}_{q})$ for $\tilde{\mathcal{Q}}$, and compute $P_{illumi}^{min}$;
\WHILE {$z^{u}/z^{l}>1+\epsilon$ and $c_{r}^{\ast}<0$}
\STATE Solve RMP and obtain its optimal result $z^{u}$ and dual optimal solution $\lambda_{j}$;
\STATE Solve PP with $\lambda_{j}$ and obtain a IS $\mathcal{I}_{q}$ with $c_{r}^{\ast}$ and the corresponding optimal $P_{DC}^{i,q,m}$;
\STATE Update $c_{r}^{\ast}$ and $\tilde{\mathcal{Q}}=\tilde{\mathcal{Q}}\bigcup\mathcal{I}_{q}$, and compute the new $P_{AC,Avg}(\mathcal{I}_{q})$ and $P_{DC}(\mathcal{I}_{q})$;
\STATE $z_{l}=z^{u}+c_{r}^{\ast}$;
\ENDWHILE
\STATE Update the link capacity in RMP by reality check;
\STATE Re-optimize RMP and obtain its optimal result $z^{u}$;
\end{algorithmic}
\end{algorithm}

\section{Practical Issues}\label{pracitcal_issues}


\subsection{Three Configurations of Light Source}
\begin{figure}
\centering
\includegraphics[width=0.48\textwidth]{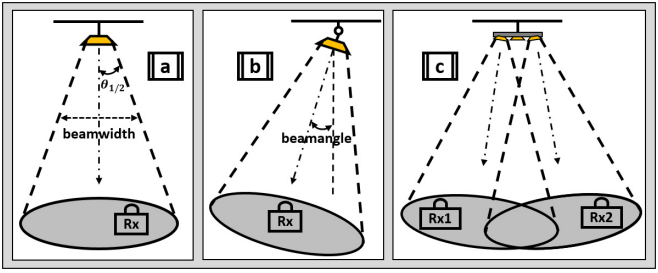}
\caption{Three different configurations: a) fixed beamangle and beamwidth; b) mechanically adjustable beamangle and beamwidth; c) electronically selectable beamangle and beamwidth}
\vspace{-5pt}
\label{fig_three_configurations}
\end{figure}

Typically, a light source operates with fixed beamangle (i.e. formed by the central luminous flux and the central vertical line) and beamwidth (i.e. semi-angle at half power $\theta_{1/2}$). This configuration is capable of efficiently providing sufficient illumination. However, its weak competitiveness manifest on the communication functionality, due to the high inter-link interference and non-uniform distribution of optical signal strength, which will lead to severe degradation of channel gain when the transceivers are not strictly aligned. To mitigate the interference and enhance the channel gain, mechanically steering the beamangle and beamwidth is introduced in \cite{tronghop2012modeling} and \cite{li2012vico}. In this paper, we further explore the impact of tuning the beamangle and beamwidth on the power consumption of multi-user VLC indoor networks. Although adjustable beamangle can be implemented by traditional {\it mechanical steering} method, the energy cost of tuning the orientation of light source is considerable \cite{nakhkoob2009multi}, and even worse, the time it takes to change the direction is inevitable, which will bring a significant negative-impact on the network delay and capacity. To circumvent these problems, a new structure of light source is proposed. Multiple chips with different beamangles and beamwidths are installed on one light source and the diversity of orientation of light source can be implemented by activating any of those chips. This configuration is motivated by the ``{\it electronic steering}" concept introduced in \cite{nakhkoob2009multi}. Since the new structure of light source needs to be specified before being evaluated by the optimization algorithm, we outline an implementable design of such light source based on the assumption of uniformly distributed UTs. In summary, three configurations\footnote{In order to distinguish the difference among three configurations, it is assumed that, in the first and the second configurations, the beamangle of all the transmitters on one VLC AP are the same.} (shown in Fig.~\ref{fig_three_configurations}) of light source are investigated: i) fixed beamangle and beamwidth; ii) mechanically adjustable beamangle and beamwidth; iii) electronically selectable beamangle and beamwidth.

As shown in Fig.~\ref{fig_three_configurations}, for Config. a, the beamwidth and beamangle are fixed regardless of the location of UT. The orientation of all VLC APs are vertical to the horizontal UTs plane. The semi-angle at half power $\theta_{1/2}$ (i.e. beamwidth) of AC and DC powered sources are the same. For Config. b, the beamwith and beamangle are tunable. For the purpose of concentrating the optical signal, the $\theta_{1/2}$ of AC powered source is set to be smaller than that of DC powered source. By adjusting the beamangle, the transmitter can directly point to the receiver. It can be seen that, for Config. b, the radiance angle of each link is always zero. For Config. c, multiple chips are installed on one VLC AP. The beamangle and beamwidth of each chip are fixed. Nevertheless, the VLC AP can selectively activate any of those chips and ``electronically steering" the beamangle and beamwidth. Notice that, since the beamangle of each chip (for Config. c) is fixed, the transmitter might not be able to directly point to the receiver, such as for Config. b. Next, we specify the new structure of light source for Config. c.

\subsection{New Structure of Light Source}

\begin{wrapfigure}{r}{0.21\textwidth}
  \vspace{-5pt}
  \begin{center}
    \includegraphics[width=0.21\textwidth]{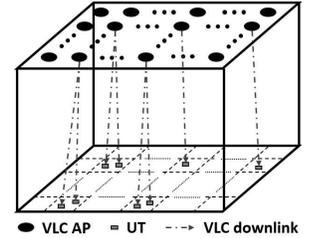}
  \end{center}
  \vspace{-10pt}
  \caption{Grid structure of multi-user VLC indoor network}
  \vspace{-5pt}
  \label{fig_grid_structure_VLC}
\end{wrapfigure}

\begin{figure}
\centering
\includegraphics[width=0.40\textwidth]{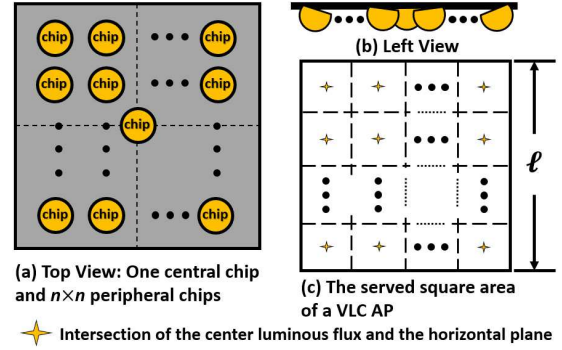}
\caption{New structure of light source}
\vspace{-5pt}
\label{fig_acr_shaped_general}
\end{figure}

As shown in Fig.~\ref{fig_grid_structure_VLC}, the VLC APs are mounted on the ceiling in grid structure. To maximize the capacity of each link, each UT\footnote{To focus on the configuration of light source and make it easier to understand, it is assumed here and also in the simulation that, each UT is equipped with only one receiver. It means that each UT can connect to only one VLC AP at a time.} chooses the nearest VLC AP to associate with, thus the served area of each VLC AP can be modeled as a square. In Fig.~\ref{fig_acr_shaped_general}, a new structure of light source is shown. It can be seen from Fig.~\ref{fig_acr_shaped_general} (a) that, one VLC AP is equipped with $(n\times n)+1$ chips, including $n\times n$ (n is a multiple of 2) peripheral chips and one central chip, which is mainly responsible for illuminating.  In Fig.~\ref{fig_acr_shaped_general} (c), a square area served by a VLC AP is equally divided into $n\times n$ small square regions. Each peripheral chip is assigned to serve one region. The star (i.e. center of coverage) in each region is the intersection of the central luminous flux of the corresponding peripheral chip and the horizontal UT plane. Each star is located at the center of each square region. Denote the side length of a square area served by a VLC AP as $l$, the maximum distance $d_{max}$ between a UT and its corresponding center of coverage satisfy $d_{max}\leq \frac{\sqrt{2}l}{2n}$.

\subsection{Constructing conflict graph}
In this section, we also introduce an effective approach to construct the conflict graph for multi-user VLC indoor networks. Traditionally, in Protocol Model, whether or not a transmission is being interfered by another transmission is determined by the distance between the receiver and the non-intended transmitter \cite{gupta2000capacity,shi2013bridging,shi2007optimal}. In VLC network, this standard is not always applicable, since the optical signal propagation may not be isotropic from the perspective of UTs plane. In \cite {tao2015scheduling}, the orientation of each VLC AP is fixed and vertical to the horizontal UTs plane, such that the conflict graph, constructed by using the interference range \cite{gupta2000capacity} as in RF networks, is valid. However, if the orientation of VLC APs is tunable, then interference range is not a reliable measure to accurately model the level of interference. As such, we propose a new criterion, according to which, if one of the pairwise signal-to-interference ratio (SIR) measurement of two links is lower than a threshold $SIR_{th}$, those two links can not be scheduled in the same IS.


Under Protocol Model, we define a maximum SIR threshold as $SIR_{th}^{max}$ such that if $SIR_{th}\geq SIR_{th}^{max}$, then when any link is active, other links on the same channel can not be active. A minimum SIR threshold $SIR_{th}^{min}$ is 1, since if $SIR_{th}<1$, the interference signal will be stronger than the transmission signal. For a wireless network with high traffic load, there could exist an upper bound $SIR_{th}^{U}$ for SIR threshold such that if $SIR_{th}>SIR_{th}^{U}$, the Protocol Model solution is infeasible (i.e. $\sum_{1\leq q\leq|\mathcal{Q}|}\omega{q}>1$). For a dense wireless network, there could exist a lower bound $SIR_{th}^{L}$ such that if $SIR_{th}<SIR_{th}^{L}$, even though the Protocol Model solution is feasible (i.e. $\sum_{1\leq q\leq|\mathcal{Q}|}\omega{q}\leq1$), the reality check result is infeasible. Therefore, we need to select the $SIR_{th}$ within the range $[SIR_{th}^{L}, SIR_{th}^{U}]$. In the next section, we study the $SIR_{th}^{U}$ and the $SIR_{th}^{L}$ for a VLC indoor network with three different configurations of light source through simulations.


\section{Numerical Results}\label{numerical_analysis}
In this section, we conduct extensive simulations to evaluate our proposed algorithm integrated with the practical issues. The simulations are conducted under Matlab R2013b and CPLEX 12.6.1 \cite{cplex2014v12} on a computer with 2.0 GHz and 4 GB RAM. We study the cost of solving the MP under different values of $\epsilon$. Our proposed algorithm is compared with two other recent schemes \cite{li2012vico,tao2015scheduling} in terms of power consumption and running time. In \cite{li2012vico}, the VICO framework does not consider the illumination constraints in the algorithm. Let $P_{total}$ represent the total power consumption of a given scheme, since the illumination is always needed, the power consumption evaluated in the numerical analysis is $P_{total}-P_{illumi}^{min}$.

\begin{figure}
\centering
\includegraphics[width=0.30\textwidth]{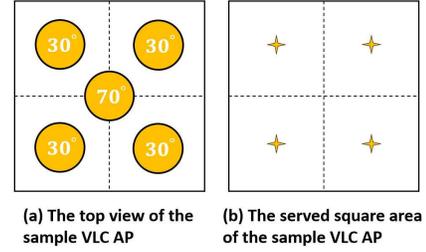}
\caption{A sample new structure light source}
\label{fig_acr_shaped_arch}
\end{figure}

The room size is 6.0 m $\times$ 6.0 m $\times$ 3.0 m. The height of desk is 0.8 m and the UTs are randomly distributed on the desk according to uniform distribution. There are 36 (6$\times$6) VLC APs in grid structure installed on the ceiling and each VLC AP is equipped with one chip\footnote{It means that each VLC AP can serve only one UT at a time.} (i.e. transmiter) filled with 625 (25$\times$25) white light LEDs. Note that, for the new structure of light source (Fig.~\ref{fig_acr_shaped_arch}), each VLC AP is equipped with 5 such chips\footnote{In order to fairly evaluate the three configurations, for Config. c, it is assumed that only one of the peripheral chips can be active for data transmission at a time, and the total optical power generated from one VLC AP can not exceed the $P_{max}$ of a single chip.}. The distances between two neighboring VLC APs and two neighboring LEDs are 1 m and 1 cm, respectively. The maximum transmitted optical power of each LED is 20 mW. The value of $P_{AC}$ of each VLC AP is set to 0.1 W. $P_{AC,Avg}$ is set to be half of $P_{AC}$. The channel bandwidth of each VLC link is 100 MHz and we assume all the VLC APs use the same channel in the simulation. The constant Gaussian noise is calculated from the parameters in \cite {komine2004fundamental} and set to be 4.7$\times$10$^{-14}$ A$^{2}$. The receiver parameters (i.e. FOV of receiver, detector area of a photodiode, gain of optical filter, refractive index of lens and O/E conversion efficiency) are the same with those in \cite {komine2004fundamental}. The required illuminance range is 300-500 lux. For Config. a, the semi-angle at half power $\theta_{1/2}$ of AC and DC powered sources are both set to 70$^\circ$. For Config. b, $\theta_{1/2}$ of AC powered source is set to 30$^\circ$, while that of DC powered source is set to 70$^\circ$. For Config. c, as shown in Fig.~\ref{fig_acr_shaped_arch}, $\theta_{1/2}$ of the central chip is set to 70$^\circ$, and $\theta_{1/2}$ of the peripheral chips are all set to 30$^\circ$. The wall plug efficiency factors $\eta_{AC}$ and $\eta_{DC}$ for all the VLC transmitters are set to 0.02 and 0.1, respectively.

\begin{figure*}
  \centering
  \begin{minipage}[t]{.329\linewidth}
    \includegraphics[width=1.0\textwidth]{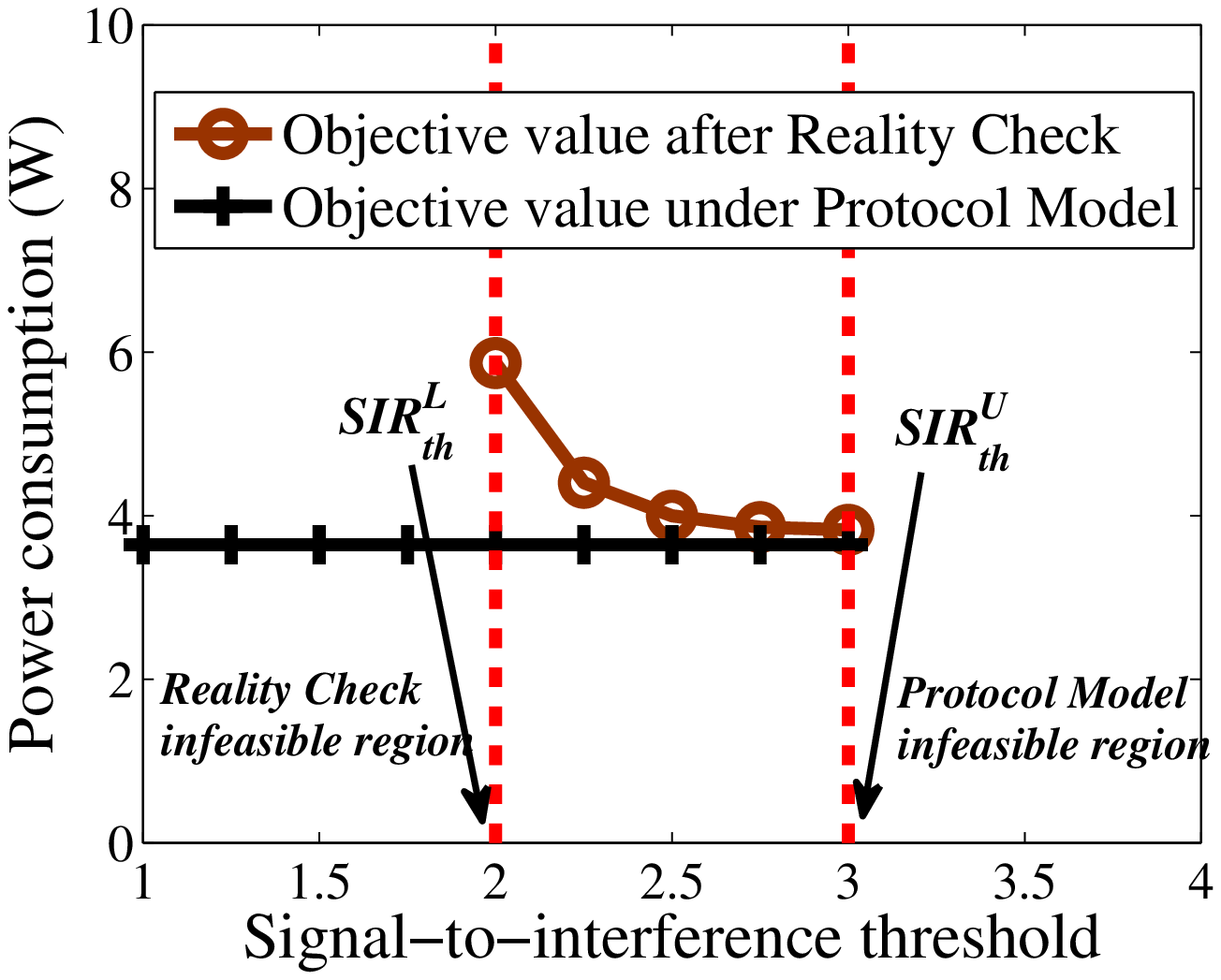}
  \vspace{-10pt}
  \caption{Protocol Model solutions and corresponding reality check results}
  \vspace{-2pt}
  \label{fig_SIR_sample}
  \end{minipage}
  \begin{minipage}[t]{.329\linewidth}
   \includegraphics[width=1.0\textwidth]{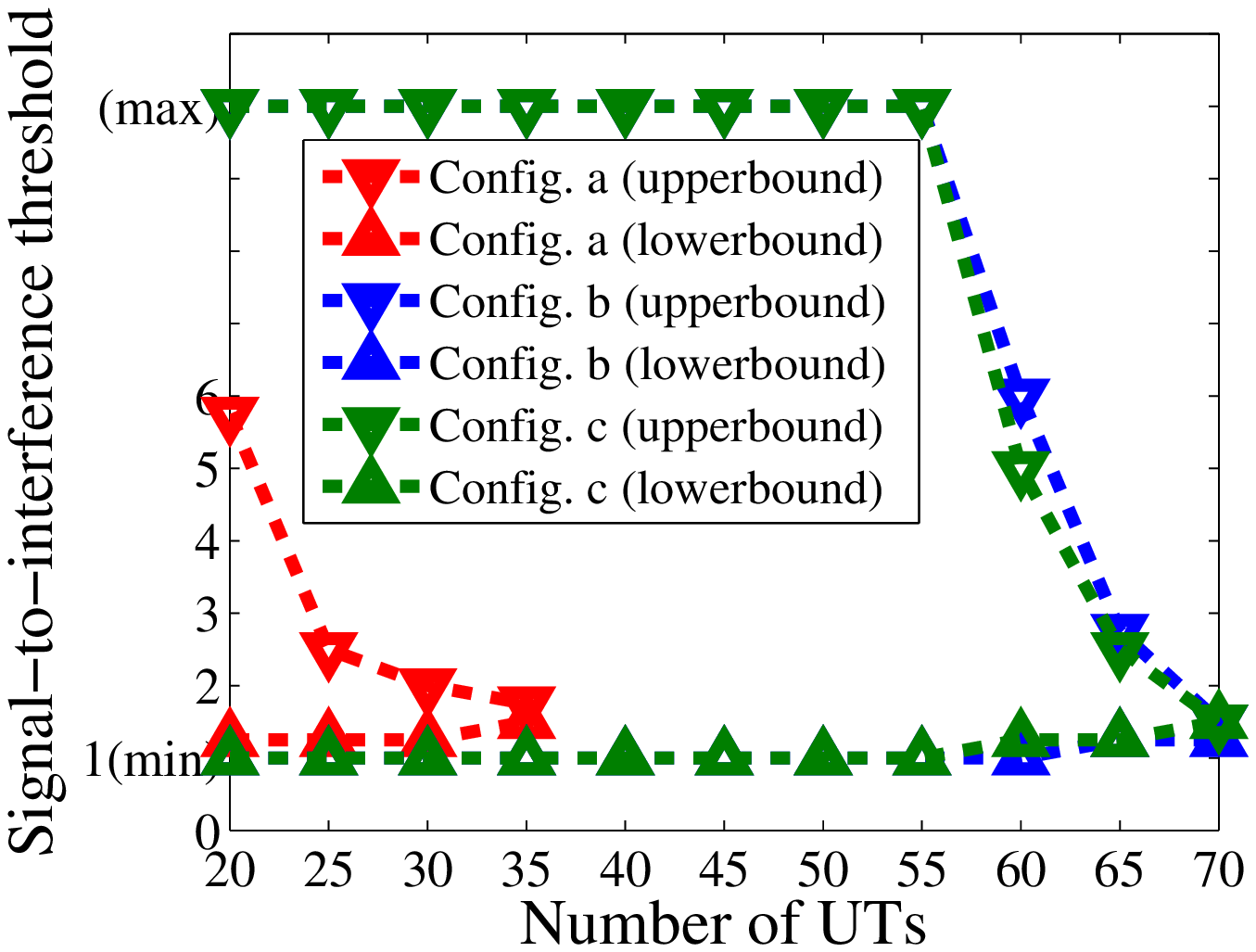}
  \vspace{-10pt}
  \caption{$SIR_{th}^{L}$ and $SIR_{th}^{U}$ for three configurations}
  \vspace{-2pt}
  \label{fig_effective_SIR}
  \end{minipage}
  \begin{minipage}[t]{.329\linewidth}
    \includegraphics[width=1.0\textwidth]{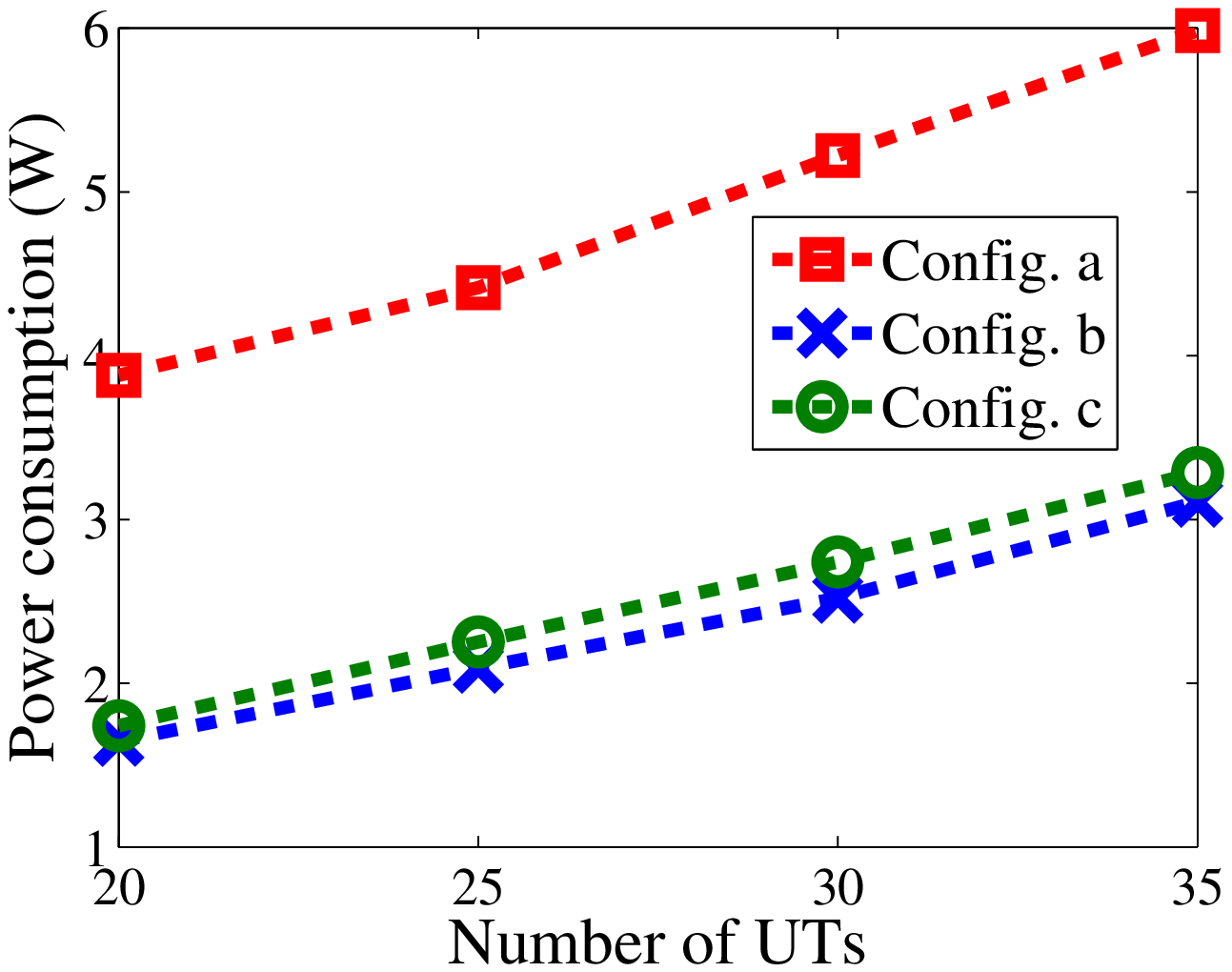}
  \vspace{-10pt}
  \caption{Real power consumption under different number of UTs for three configurations}
  \vspace{-2pt}
  \label{fig_user_three_cases}
  \end{minipage}
\end{figure*}

\subsection{Cost of solving MP}
For a 30-UTs VLC indoor network, Table~\ref{table_complexity} shows the number of iterations and the total running time needed to achieve an $\epsilon$-bounded solution. It can be seen that, it takes 14 iterations and 9.33 seconds to obtain 1\%-bounded results, and 44 iterations and 17.84 seconds to get a near-optimal result (i.e. $\epsilon\approx0$). For the following numerical results, we use $\epsilon=0.01$.

\begin{table}
\centering
\caption{Iteration Number and Running Time for an $\epsilon$-bounded solution}
\begin{tabular}{|c|c|c|}
\hline
$\epsilon$&Number of Iterations&Running Time (s)\\\hline
0.01&14&9.33\\
0.005&22&11.56\\
1$\times$10$^{-14}$&44&17.84\\
\hline
\end{tabular}
\label{table_complexity}
\end{table}

\subsection{Constructing Conflict Graph}
For a 30-UTs VLC indoor network under Config. a, the traffic demand of each UT is 20 Mbps, we show the Protocol Model solutions and the corresponding reality check results in Fig.~\ref{fig_SIR_sample}. The upper bound for SIR threshold $SIR_{th}^{U}$ is 3, and if $SIR_{th}>3$, the Protocol Model solution will be infeasible. The lower bound for SIR threshold $SIR_{th}^{L}$ is 2, and if $SIR_{th}<2$, the reality check solution will be infeasible. When $2\leq SIR_{th}\leq3$, although the Protocol Model solution is almost the same, a lower $SIR_{th}$ will lead to a higher reality check result, which is due to neglecting the non-zero interference in Protocol Model. Thus, choosing $SIR_{th}^{U}$ as the SIR threshold will minimize the total power consumption while satisfying the traffic demands.

In Fig.~\ref{fig_effective_SIR}, we show the $SIR_{th}^{U}$ and the $SIR_{th}^{L}$ for a VLC indoor network under the three configurations. As the number of UTs increases, the $SIR_{th}^{U}$ will decrease and the $SIR_{th}^{L}$ will increase. And when the $SIR_{th}^{U}$ and the $SIR_{th}^{L}$ converge to one value, this indicates that the traffic load reaches up to the system capacity. From Fig.~\ref{fig_effective_SIR}, we can observe that, the system capacity of Config. b and Config. c is around double that of Config. a. And the feasible region of Config. b and Config. c is larger than that of Config. a, which shows the superiority of the new structure of light source (i.e. Config.~c).

\subsection{Power Consumption Evaluation}
In the following, we evaluate the objective value (i.e. power consumption) for the three configurations, in terms of different number of UTs. From Fig.~\ref{fig_user_three_cases} , it can be observed that, the real power consumption (after reality check) for Config. a is more than double that for Config. b and Config. c. More to the point, the real power consumption for Config. c, which is much more practical than Config. b, is almost the same as that for Config. b. In summary, the new structure of light source (i.e. Config. c) is a practical design and capable to provide excellent power-saving performance.

\begin{figure*}
  \centering
  \begin{minipage}[t]{.327\linewidth}
    \includegraphics[width=1.0\textwidth]{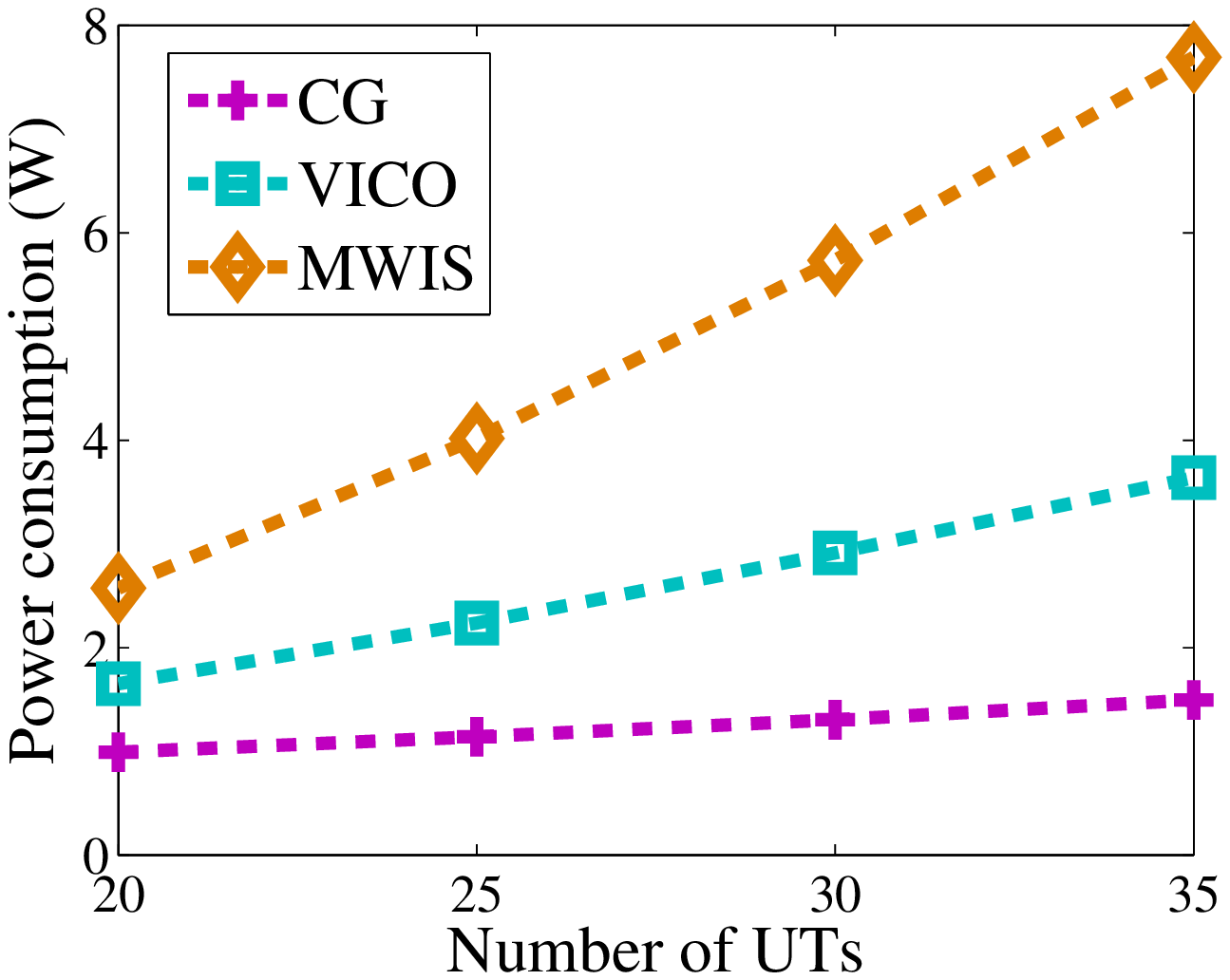}
  \vspace{-10pt}
  \caption{Real power consumption under different number of UTs for three algorithms}
  \vspace{-2pt}
  \label{fig_user_three_algorithms}
  \end{minipage}
  \begin{minipage}[t]{.327\linewidth}
   \includegraphics[width=1.0\textwidth]{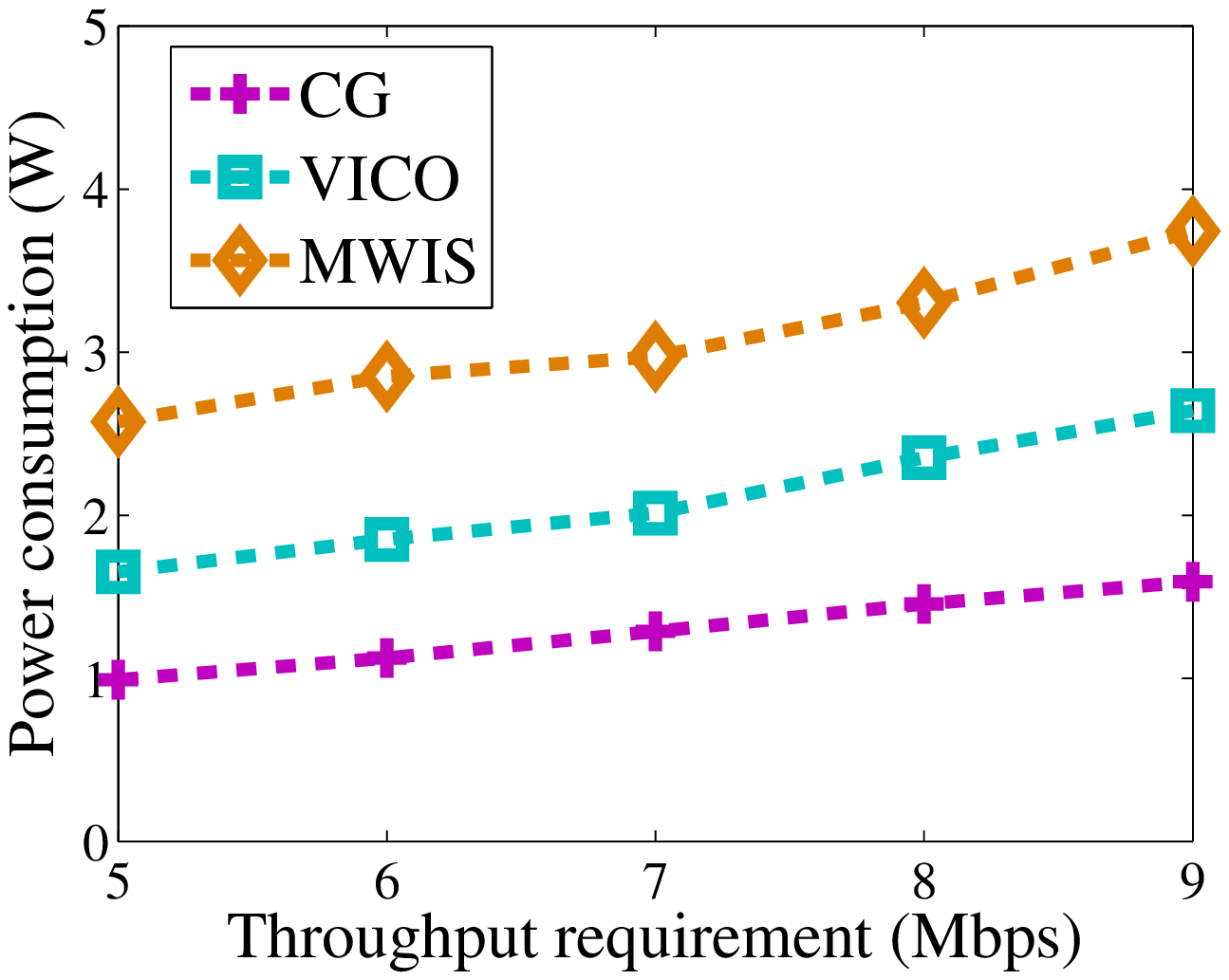}
  \vspace{-10pt}
  \caption{Real power consumption under different throughput requirement for three algorithms}
  \vspace{-2pt}
  \label{fig_throughput_three_algorithms}
  \end{minipage}
  \begin{minipage}[t]{.327\linewidth}
    \includegraphics[width=1.0\textwidth]{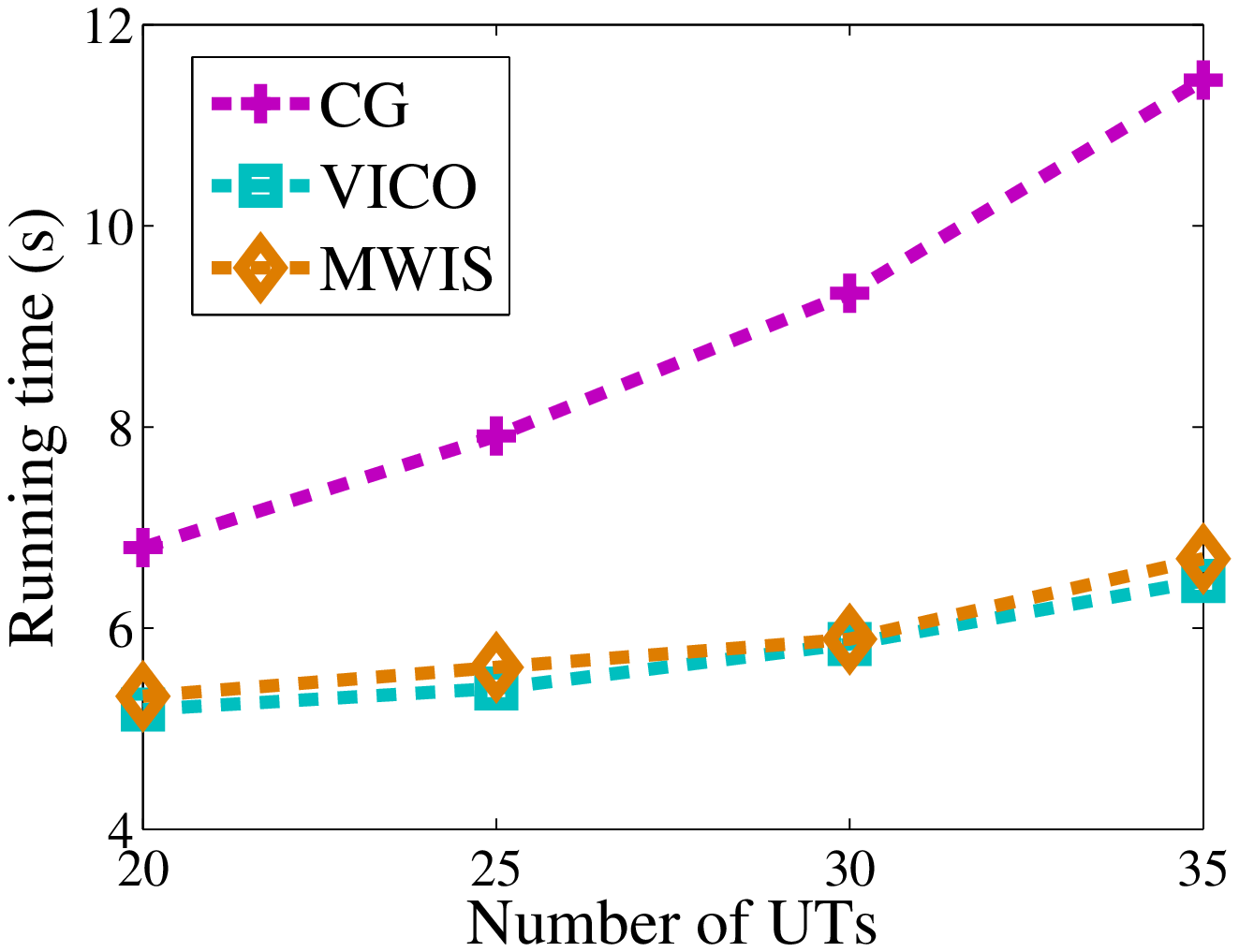}
  \vspace{-10pt}
  \caption{Algorithm running time under different number of UTs for three algorithms}
  \vspace{-2pt}
  \label{fig_complexity_three_algorithms}
  \end{minipage}
\end{figure*}

%

We also compare the $\epsilon$-bounded solution obtained by column generation (CG) with the results obtained by a random link scheduling introduced in VICO \cite{li2012vico} and a link scheduling based on maximum weighted independent set (MWIS) introduced in \cite{tao2015scheduling}. The real power consumption under the three algorithms are simulated in terms of different number of UTs and different throughput requirement of each UT. As shown in Fig.~\ref{fig_user_three_algorithms}, the traffic demand of each UT is 5 Mbps, the real power consumption under VICO and MWIS becomes much higher than that under CG, as the number of UTs increases. For a 35-UTs network, the CG algorithm cuts the power consumption of VICO by 60\%. Although the MWIS algorithm can achieve a better objective value under Protocol Model than random scheduling, after reality check its power consumption is much worse than that of VICO due to neglecting the non-zero interference from other links. The results shown in Fig.~\ref{fig_throughput_three_algorithms} are consistent with the above analysis. For a 20-UTs VLC network, as the throughput requirement of each UT increases, there is almost no changes in the gaps between the real power consumption obtained by three algorithms.


%

\subsection{Illumination Evaluation}
\begin{figure}
\centering
\includegraphics[width=0.48\textwidth]{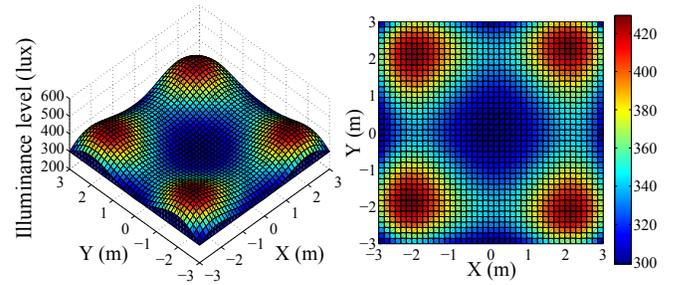}
\caption{A sample distribution of illuminance created by our proposed algorithm with the entire horizontal space illumination constraints}
\vspace{-7pt}
\label{fig_CG_illumination}
\end{figure}

\begin{figure}
\centering
\includegraphics[width=0.48\textwidth]{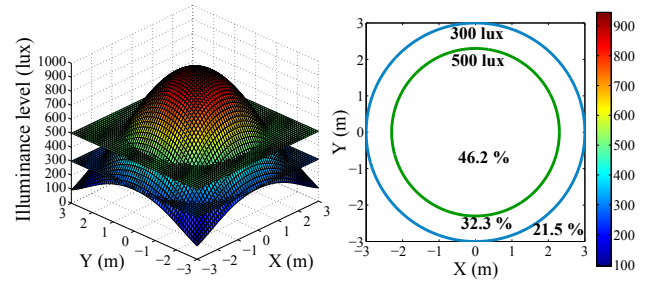}
\caption{A sample distribution of illuminance under the VICO framework}
\vspace{-7pt}
\label{fig_VICO_illumination}
\end{figure}

Regarding the illuminance satisfaction, in our proposed algorithm, since the illuminance level for the entire horizontal space is conditioned to be within the specified range (300-500 lux), the illuminance distribution always meets the requirements. In Fig.~\ref{fig_CG_illumination}, we show a sample distribution of illuminance under our proposed algorithm. We also show a sample illuminance distribution under VICO framework in Fig.~\ref{fig_VICO_illumination}. Around 70\% of the horizontal space do not meet the illumination requirements, because VICO framework does not take the illumination constraints into account when performing the optimization algorithm.

\subsection{Cost of three algorithms}
The running time costs by minimizing the power consumption under three algorithms are shown in Fig.~\ref{fig_complexity_three_algorithms}. We can observe that, the CG algorithm costs around 5.2 seconds more than that of the random link scheduling under VICO. Nevertheless, the optimization algorithm can be run with one-time cost in the centralized controller and no extra operations are needed until the network environment changes. This extra overhead is acceptable, given that our algorithm can achieve a 60\% power-saving performance while satisfying the illumination requirements.

\section{Related Work}\label{related_work}
Many efforts have been paid on the VLC indoor network, however, most of them do not provide comprehensive consideration on the illumination and communication, or only focus on limited aspects. Even though one of them \cite{li2012vico} proposes a framework for configuring the VLC indoor networks with adjustable LEDs’ beamangle and beamwidth, the link scheduling algorithm has no power efficiency guarantees and the framework relies on a simple impractical assumption to measure the illumination using the average SNR distribution.

The LED angle tuning issue is also studied in \cite{tronghop2012modeling}, which only presents the heuristic simulation results of the average SNR distribution without considering illumination. For the cooperation among VLC cells, the work in \cite{li2015cooperative} presents an elaborative study on the bandwidth efficiency of four different VLC cell formations. However, in the cooperative load balancing approach, the users’ traffic demand is not concerned. The merging VLC cell approach is also studied in \cite{tao2015scheduling}, which has not taken the illumination into account. In \cite{wu2014cellular}, an optimal Lambertian order algorithm is proposed. To maximize the VLC cell boundary signal strength, an optimal Lambertian order can be found by calculating the first order derivative of the VLC path loss model. Whereas the illumination and multiple users scenario are not considered in this work. The work in \cite{rahaim2013sinr} introduces the concept of DC and AC optical power without considering overall power consumption optimization and the distribution of illumination. In \cite{din2014energy}, the authors investigate the energy-efficiency for only optical power with brightness control and data transmission in VLC networks. Three power levels of the sub-carrier pulse position modulation scheme are utilized as the variables to perform the power consumption optimization. Nonetheless, in their optimization constraints, the brightness control is only applied to the location of user, which is not practicable. A multi-transceiver optical wireless spherical structure is proposed in \cite{nakhkoob2009multi}. The spherical optical antenna is tessellated with multiple LEDs, to enable the beamangle diversity. However, this design does not carefully consider the illumination functionality of LEDs. The transverse line-of-sight of the spherical optical antenna may produce glaring in some specific scenarios. Several brightness control methods have been introduced in \cite{sugiyama2007brightness}. The PWM and DC bias are two predominant approaches for dimming control. A room division multiplexing-based VLC network is proposed in \cite{huang2013design}, while in our work, the concept of space division multiplex is manifested by adjusting the Lambertian order.

\section{Conclusion}\label{conclusion}
In this paper, we investigate the problem of minimizing the total power consumption of a general multi-user VLC indoor network while satisfying the traffic demands and providing acceptable level of illumination. A column generation based $\epsilon$-bounded approximation algorithm with reality check is proposed. Regarding the light source, three configurations are considered and one of them (i.e. a new structure of light source) is proposed. For constructing the conflict graph, an effective range of SIR threshold is evaluated by extensive simulation results. The power consumption of the three configurations are evaluated by simulations and the results reveal that the new structure light source is practical and provides power-efficient solutions. Compared to two other VLC link scheduling algorithms, our proposed algorithm can achieve a better performance of power consumption, especially in crowded scenarios, while satisfying the illumination requirements on the entire horizontal plane.

\section*{Acknowledgment}
This work was supported in part by the NSF grant ECCS-1331018.

\bibliographystyle{IEEEtran}
\bibliography{Infocom_paper}

\section*{Author's Biographies}
\textbf{Sihua Shao} received his B.S. degree in electrical and information engineering from South China University of Technology in 2011, and his M.S. degree in electrical and information engineering from Hong Kong Polytechnic University in 2012. He is currently a Ph.D student in the Department of Electrical and Computer Engineering at New Jersey Institute of Technology. His current research interests include wireless communication, visible light communication and heterogeneous network. He is a student member of the IEEE.

\textbf{Abdallah Khreishah} received his Ph.D and M.S. degrees in Electrical and Computer Engineering from Purdue University in 2010 and 2006, respectively. Prior to that, he received his B.S. degree with honors from Jordan University of Science and Technology in 2004. During the last year of his Ph.D, he worked with NEESCOM. He is currently an Assistant Professor in the ECE department of New Jersey Institute of Technology. His research spans the areas of green communications and networking, visible-light communications, wireless networks, and network security. His research projects are funded by the National Science Foundation, New Jersey Department of Transportation, and the UAE Research Foundation.

\textbf{Issa Khalil} received the B.Sc. and the M.Sc. degrees from Jordan University of Science and Technology in 1994 and 1996 and the PhD degree from Purdue University, USA in 2006, all in Computer Engineering. Immediately thereafter he worked as a post-doctoral researcher in the Dependable Computing Systems Lab of Purdue University until he joined the College of Information Technology (CIT) of the United Arab Emirates University (UAEU) in August 2007. In September 2011 Khalil was promoted to associate professor and served as the department chair of the Information Security Department in CIT. In June 2013, Khalil joined the Cyber Security Group in Qatar Computing Research Institute (QCRI), a member of Qatar Foundation, as a Senior Scientist. Khalil’s research interests span the areas of wireless and wireline network security and privacy. He is especially interested in cloud security, botnet detection and takedown, and security data analytics. Dr. Khalil served as the technical program co-chair of the 6th International Conference on Innovations in Information Technology and was appointed as a Technical Program Committee member and reviewer for many international conferences and journals. In June 2011, Khalil was granted the CIT outstanding professor award for outstanding performance in research, teaching, and service.

\end{document}